\documentclass[12pt]{article}
\begin{document}

\begin{titlepage}

\begin{flushright}
TIT/HEP-515\\
{\tt hep-th/0312149}\\
December, 2003\\
\end{flushright}
\vskip 1cm

\begin{center}
{\Large \bf Rolling Tachyon with Electromagnetic Field\\
\vskip 3mm
in Linear Dilaton Background} \\
\vskip 2cm
{\large \bf Kenji~Nagami} \\
\vskip 0.5cm
{\large \it Department of Physics, Tokyo Institute of Technology,\\
Meguro, Tokyo 152-8551, Japan} \\
\vskip 2mm
{\small {\tt nagami@th.phys.titech.ac.jp}} \\
\end{center}

\vskip 1cm

\begin{abstract}
Rolling tachyon in linear dilaton background is examined by using an effective field theory with gauge field on an unstable D-brane in bosonic string theory.
Several solutions are identified with tachyon matter equipped with constant electromagnetic field in linear dilaton background.
The time evolution of effective coupling and the nature of propagating fluctuation modes around the tachyon matter are also studied.
\end{abstract}

\end{titlepage}

\normalsize
\newpage
\setcounter{page}{1}
%%%%%%%%%%%%%%%%%%%%%%%%%%%%%%%%%%%%%%%%%%%%%%%%%%%%%%%%%%%%%
%%%%%%%%%%%%%%%%%%%%%%%%%%%%%%%%%%%%%%%%%%%%%%%%%%%%%%%%%%%%%
%%%%%%%%%%%%%%%%%%%%%%%%%%%%%%%%%%%%%%%%%%%%%%%%%%%%%%%%%%%%%
\section{Introduction}
The decay of unstable D-brane can be described by tachyon condensation.
The time dependent process of tachyon condensation called rolling tachyon has recently received much attention.
It is found that the tachyon condensation leaves behind some objects equipped with the same energy density as the original unstable D-brane tension and the vanishing pressure~\cite{Sen:2002nu}.
These objects are called tachyon matter or tachyon dust.

The degree of freedom of perturbative open string disappears under tachyon condensation.
On the other hand, electric flux on D-brane world volume is generically identified with fundamental string charge for an anti-symmetric 2-tensor called the Kalb-Ramond field, then the electric flux must be conserved under tachyon condensation.
Accordingly, only the confined electric flux can be allowed to remain and play a role as the carrier of fundamental string charge.
Indeed, in super string theory, the confinement of unbroken gauge field is well described in~\cite{Yi:1999hd,Bergman:2000xf}. It is expected that the electric field is to be confined and identified with fundamental string in bosonic string theory too~\cite{Rey:2003zj,Sen:2003bc}.

The rolling tachyon accompanied with a constant electric field is examined by using the effective field theories~\cite{Gibbons:2002tv} and the boundary CFT~\cite{Mukhopadhyay:2002en,Rey:2003xs}. It is found that the energy density and the pressure along the constant electric field become certain finite constants and pressure along directions transverse to the electric field becomes zero. This solution can be identified with tachyon matter with a constant electric field. Moreover it is observed that fluctuation modes around this solution can propagate along the electric field at speed proportional to the field strength~\cite{Gibbons:2002tv}.
The nature of propagation of fluctuation modes is likely to suggest a significant implication of fundamental string formation and electric flux confinement.

The rolling tachyon in linear dilaton background is examined by using the timelike boundary Liouville theory.
It is shown that there is a solution to be identified with tachyon matter.
Moreover it is observed that the tachyon matter slides towards a weak coupling region, assuming the normalizability of a local operator denoting tachyon field~\cite{Karczmarek:2003xm}.
Accordingly, it is intriguing to construct the tachyon matter solution in linear dilaton background by using an effective field theory, and examine whether a tendency for tachyon matter to slide towards a certain coupling region can be shown in the effective field theory.

In this paper, we will consider the rolling tachyon on an unstable D$p$-brane in a linear dilaton background in bosonic string theory, by using an effective field theory with gauge field.
In the next section, we begin with a brief review of a useful prescription to analyze tachyon condensation with gauge field~\cite{Gibbons:2002tv,Gibbons:2000hf}.
Initially, we examine the rolling tachyon without gauge field in linear dilaton background, and obtain asymptotic solutions.
By studying the asymptotic behavior of energy momentum tensor, we identify a solution to be tachyon matter in linear dilaton background.
We are also lead to an unexpected solution as obtained in~\cite{Kluson:2003sr}.
Subsequently, we study the rolling tachyon accompanied with a constant electromagnetic field in linear dilaton background.
The time evolution of effective coupling and the propagation of fluctuation modes around the tachyon matter solutions are also examined.
We work with the weak string coupling limit.
%%%%%%%%%%%%%%%%%%%%%%%%%%%%%%%%%%%%
%%%%%%%%%%%%%%%%%%%%%%%%%%%%%%%%%%%%
\section{rolling tachyon in linear dilaton background}
The dynamics of tachyon $T(x)$ and gauge fields $A_{\mu}(x)$ on an unstable D$p$-brane with dilaton $\Phi(X(x))$ would be described by a world volume effective action given by
\begin{equation}
S=-\tau_p\int d^{p+1}x\, e^{-\Phi} V(T) \sqrt{-\det (\eta_{\mu\nu}+F_{\mu\nu}+\partial_{\mu}T \partial_{\nu}T)}.
\label{action}
\end{equation}
This is a simple generalization of the effective action proposed in~\cite{Garousi:2000tr,Bergshoeff:2000dq} to the case involving the dilaton.
We will consider a linear dilaton background $\Phi=V_{\mu}X^{\mu}$ that has non-vanishing components only along the D-brane world volume, and use the static gauge $X^{\mu}=x^{\mu}\hspace{3mm}(\mu=0,1,\cdots,p)$ where $X^{\mu}$ denotes target space coordinate and $x^{\mu}$ denotes world volume one.
The target space is flat with metric signature $(-+\cdots +)$, and $\tau_p$ denotes the D$p$-brane tension. $\alpha'=1$ unit is used in this paper.
The tachyon enters into the action as if it is one of transverse coordinates on D-brane, which is related to gauge field under T-duality.
Indeed, it is possible to treat the tachyon and gauge field on an equal footing except for the presence of potential.
The following relation on matrix determinant,
\begin{equation}
\det(P)\det(Q+RP^{-1}R^{T})=\det(Q)\det(P+R^{T}Q^{-1}R)=
\det\left(
\begin{array}{cc}
P&-R^{T} \\
R&Q
\end{array} \right),
\end{equation}
with identification as $P=1, Q=(\eta+F)_{\mu\nu}, R=\partial_{\mu} T$, allows the action (\ref{action}) to be expressed by
\begin{equation}
S=-\tau_p \int d^{p+1}x\, e^{-\Phi} V(T) \sqrt{-\det (\eta_{MN}+F_{MN})}.
\label{action2}
\end{equation}
The tachyon field $T$ is incorporated with the gage field $A_{\mu}$ to define an extended gauge field as in~\cite{Gibbons:2002tv,Gibbons:2000hf},
\begin{equation}
A_M= \left\{
\begin{array}{ll}
T,       & \hspace*{3mm}(M=-1)  \nonumber \\ 
A_{\mu}, & \hspace*{3mm}(M=\mu),\nonumber
\end{array} \right.
\end{equation}
where the world volume coordinate $x^{\mu}\hspace{2mm}(\mu=0,1,\cdots,p)$ is also extended to $x^{M}\hspace{2mm}(M=-1,0,1,\cdots,p)$ in this order.
The extended flat metric $\eta_{MN}$ is understood to have signature $(+-+\cdots+)$.
A relevant extended field strength is determined by
\begin{equation}
F_{MN}= \left\{
\begin{array}{ll}
F_{\mu\nu} & \hspace*{3mm}(M=\mu, N=\nu)\\
\partial_{\mu}T & \hspace*{3mm}(M=\mu, N=-1),
\end{array} \right.
\end{equation}
with $\partial_{-1}\equiv 0$.
This extended field strength is anti-symmetric and obeys the Jacobi identity.
The equations of motions of the action (\ref{action2}) are given by
\begin{equation}
V'(T)\left(\theta^{MN}\partial_N T-\delta^{M,-1}\right)+V(T)G^{MN}\partial_K F_{LN}G^{KL}-V(T)\theta^{MN}V_N=0,
\label{eom}
\end{equation}
with $V_{-1} \equiv 0$. The symmetric part and the anti-symmetric part of inverse matrix of $A_{MN}\equiv (\eta+F)_{MN}$ are denoted by
\begin{equation}
G^{MN}\equiv\left(\frac{1}{\eta+F} \right)^{MN}_{\mbox{sym.}},\hspace{5mm}
\theta^{MN}\equiv\left(\frac{1}{\eta+F} \right)^{MN}_{\mbox{anti-sym.}},
\label{G-theta}
\end{equation}
respectively.
The tachyon potential $V(T)$ can be taken to be such that it has a local maximum $1$ at $T=0$ and a local minimum $0$ at $T=\infty$.
The asymptotic profile of tachyon potential for very large value of $T$ suffices to construct asymptotic solutions.
Assuming the tachyon potential $V(T)=1/\cosh(T/2)$ proposed in~\cite{Lambert:2003zr,Kutasov:2003er}, one can take advantage of the asymptotic form of it given by
\begin{equation}
V(T)=e^{-\frac{1}{2}T}.
\end{equation}
We shall examine the rolling tachyon by using this potential.
%%%%%%%%%%%%%%%%%%%%%%%%%%%%%%%%%%%%%%
%%%%%%%%%%%%%%%%%%%%%%%%%%%%%%%%%%%%%%
\subsection{rolling tachyon with constant block diagonal field strength}
We will look for homogeneous rolling tachyon solutions $T=T(x^0)$ with $F_{\mu\nu}$ constant.
Lorentz rotation can be utilized for this type of solution to simplify $F_{\mu\nu}$ so that it has vanishing components except for $F_{\mu,\mu+1}\;(\mu=0,1,\cdots,p-1)$ and their transpositions.
For simplicity, we begin with a block diagonal form of $F_{\mu\nu}$ such that $F_{2i,2i+1}=-F_{2i+1,2i}=B_i \; (i=0,1,\cdots,n=\left[\frac{p-1}{2}\right])$ and other components are zero.
This supposed form of gauge fields can be involved with the rolling tachyon $T=T(x^0)$ to construct the extended field strength $F_{MN}$. As a result, $A_{MN}$ is represented by $A^{(0)}\oplus_{i=1}^n A^{(i)}$ with
\begin{equation}
A^{(0)}=\left(
\begin{array}{ccc}
1       & -\dot{T} & 0 \\
\dot{T} & -1      & B_0 \\
0        & -B_0    & 1
\end{array} \right), \hspace{5mm}
A^{(i)}=\left(
\begin{array}{cc}
1 & B_i \\
-B_i & 1
\end{array} \right).
\end{equation}
with $\dot{T}=\frac{dT}{dx^0},\; \det A^{(0)}=-1+B_0^2+\dot{T}^2,\; \det A^{(i)}=1+B_i^2$. It will turn out that there are solutions for which $\det A^{(0)}$ is not vanishing.
$G^{MN}$ is represented by $G^{(0)}\oplus_{i=1}^n G^{(i)}$ with
\begin{equation}
G^{(0)}=\frac{1}{\det A^{(0)}} \left(
\begin{array}{ccc}
-1+B_0^2 & 0 & -B_0\dot{T} \\
0 & 1 & 0 \\
-B_0\dot{T} & 0 & -1+\dot{T}^2
\end{array} \right), \hspace{5mm}
G^{(i)}=\frac{1}{\det A^{(i)}} \left(
\begin{array}{cc}
1&0\\
0&1
\end{array} \right).
\end{equation}
$\theta^{MN}$ is represented by $\theta^{(0)}\oplus_{i=i}^n \theta^{(i)}$ with
\begin{equation}
\theta^{(0)}=\frac{1}{\det A^{(0)}} \left(
\begin{array}{ccc}
0        & \dot{T} & 0 \\
-\dot{T} & 0       & -B_0 \\
0        & B_0     & 0
\end{array} \right), \hspace{5mm}
\theta^{(i)}=\frac{1}{\det A^{(i)}} \left(
\begin{array}{cc}
0&-B_i\\
B_i&0
\end{array} \right).
\end{equation}
With use of these matrices, the second term of the equations of motion (\ref{eom}) turns out to vanish except for $M=-1,1$,
\begin{equation}
G^{MN}\partial_K F_{LN} G^{KL}= \left\{
\begin{array}{ll}
(-1+B_0^2)\, \ddot{T}\,(\det A^{(0)})^{-2} & \hspace*{2mm}(M=-1) \\
-B_0\,\dot{T}\,\ddot{T}\,(\det A^{(0)})^{-2} & \hspace*{2mm}(M=1).
\end{array} \right.
\end{equation}
$M=-1$ component of the equations of motion (\ref{eom}) is
\begin{equation}
(1-B_0^2)+\frac{2(1-B_0^2)}{-1+B_0^2+\dot{T}^2}\ddot{T}+2\dot{T}V_0=0.
\label{eq_M=T}
\end{equation}
$M=0$ component is
\begin{equation}
B_0 V_1=0.
\label{eq_M=0}
\end{equation}
$M=1$ component is
\begin{equation}
B_0\left[\dot{T}+\frac{2\dot{T}\ddot{T}}{-1+B_0^2+\dot{T}^2}+2V_0\right]=0.
\label{eq_M=1}
\end{equation}
$M=2i,\, 2i+1 \hspace{1mm} (i=1,\cdots,n)$ components are
\begin{equation}
B_i V_{2i+1}=0, \hspace{8mm}
B_i V_{2i}=0.
\end{equation}
It appears that $B_i\;(i=1,\cdots,n)$ can acquire a non-vanishing value only for $V_{2i}=V_{2i+1}=0$. We will classify some solutions of these equations according to the value of $B_0$. \\
%%%%%%%%%%%%%%%%%%%%%%%%%%%%%%%%%%%%%%%
Firstly, let us consider the case of non-vanishing $B_0$.
The equation (\ref{eq_M=0}) shows that this solution is possible only for $V_1=0$. The equation (\ref{eq_M=T}) can incorporates with the equation (\ref{eq_M=1}) to give two equations,
\begin{eqnarray}
2\ddot{T}-1+B_0^2+\dot{T}^2&=&0, \label{eq_B0=n0} \\
V_0&=&0.
\end{eqnarray}
The last equation shows that this type of solution with $B_0$ finite is possible only for $V_0=0$.

To determine an asymptotic behavior of rolling tachyon, let us examine an supposed form of tachyon as in~\cite{Sen:2002nu,Kluson:2003sr},
\begin{equation}
\dot{T}=\beta - K e^{-\gamma x^0}+{\mathcal O}(e^{-2\gamma x^0}).
\label{ansatz}
\end{equation}
The parameters $\beta, K, \gamma$ should be determined to be positive constants so that the tachyon rolls down the potential towards the local minimum $(T=\infty)$, and the speed of tachyon $\dot{T}$ asymptotically becomes a certain constant so as to validate the description with effective action.
Let us apply this supposed form to the equations of motion and evaluate them in order by order of $e^{-\gamma x^0}$.

This supposed form of rolling tachyon leads to $\det A^{(0)}$ expressed by
\begin{equation}
\det A^{(0)}=C-2\beta K e^{-\gamma x^0}+\cdots, \hspace{7mm}
C \equiv -1+B_0^2+\beta^2.
\label{detA-diagonal}
\end{equation}
The equation (\ref{eq_B0=n0}) evaluated in order $1$ and $e^{-\gamma x^0}$ leads to two equations among parameters. They are found to have a single solution given by
\begin{equation}
\beta=\gamma=\sqrt{1-B_0^2}.
\label{solution1}
\end{equation}
Note that the leading term $C$ of $\det A^{(0)}$ in equation (\ref{detA-diagonal}) vanishes for this solution.
The $B_0$ dependence of $\beta=\sqrt{1-B_0^2}$ shows that a final speed of rolling tachyon is slowed down in the presence of electric field $B_0$. This fact is in agreement with the result obtained in~\cite{Gibbons:2002tv}, but we note that the finite $B_0$ is allowed only for $V_1=0$.

Let us study the energy momentum tensor of the theory (\ref{action2}), which is represented by
\begin{equation}
T^{\mu\nu}=-\tau_p e^{-\Phi} V(T) \sqrt{-\det A_{MN}}\; G^{\mu\nu}, \hspace{5mm}
(\mu,\nu=0,1,\cdots,p).
\end{equation}
The solution (\ref{solution1}) leads to the block diagonal energy momentum tensor expressed by $T^{(0)}\oplus_{j=1}^n T^{(j)}$ with
\begin{eqnarray}
T^{(0)}&=&\tau_p \frac{1}{\sqrt{2\beta K}} \prod_{i=1}^n (1+B_i^2)^{1/2}
\left(
\begin{array}{cc}
1 &0 \\
0 &-B_0^2
\end{array} \right)+\cdots ,
\label{T0_B0_nonzero}\\
T^{(j)}&=&-\tau_p \sqrt{2\beta K} \prod_{i=1, i\neq j}^n (1+B_i^2)^{1/2}
(1+B_j^2)^{-1/2}
\left(
\begin{array}{cc}
1 & 0 \\
0 & 1
\end{array} \right)e^{-\gamma x^0}+\cdots.
\label{Tj_B0_nonzero}
\end{eqnarray}
The case with all $V_{\mu}$ vanishing is displayed here, but the case with $V_{2i}$ or $V_{2i+1}$ non-vanishing is readily given by replacing $(1+B_i^2)^{1/2}$ with $e^{-V_{2i}x^{2i}-V_{2i+1}x^{2i+1}}$.
In the following analysis, the case with vanishing $V_{\mu}\hspace{2mm}(\mu=2,\cdots,p)$ are mainly examined.

It is observed that the energy density $T^{00}$ remains a finite positive constant in the infinite future that is proportional to the D-brane tension $\tau_p$. The pressure becomes zero to all directions except for $x^1$. This direction is what the constant electric field is pointed to, and the pressure $T^{11}$ remains a negative constant determined by the electric field strength.
These are the suitable properties for tachyon matter with electric field~\cite{Mukhopadhyay:2002en,Rey:2003xs}.

The action includes the field strength $F_{\mu\nu}$ in a gauge invariant combination $(F_{\mu\nu}+B_{\mu\nu})$, where $B_{\mu\nu}$ denotes the pull-back of the Kalb-Ramond field $B_{MN}$ onto D-brane world volume. The source for the Kalb-Ramond field is determined by functional derivative of the static gauge action with respect to $F_{\mu\nu}$. It is expressed by
\begin{equation}
S^{\mu\nu}=\tau_p\, e^{-\Phi} V(T) \sqrt{-\det A_{MN}}\> \theta^{\mu\nu}.
\end{equation}
The solution (\ref{solution1}) leads to the block diagonal $S^{\mu\nu}$ expressed by $S^{(0)}\oplus_{j=1}^n S^{(j)}$ with
\begin{eqnarray}
S^{(0)}&=&\tau_p \frac{1}{\sqrt{2\beta K}} \prod_{i=1}^n (1+B_i^2)^{1/2}
\left(
\begin{array}{cc}
0   & -B_0 \\
B_0 & 0
\end{array} \right)+\cdots ,
\label{S0_B0_nonzero}\\
S^{(j)}&=&-\tau_p \sqrt{2\beta K} \prod_{i=1, i\neq j}^n (1+B_i^2)^{1/2}
(1+B_j^2)^{-1/2}
\left(
\begin{array}{cc}
0 & -B_j \\
B_j & 0
\end{array} \right)e^{-\gamma x^0}+\cdots .
\label{Sj_B0_nonzero}
\end{eqnarray}
Note that the fundamental string charge density pointing to $x^1$ direction, $S^{01}$, remains finite in the infinite future.
\\
%%%%%%%%%%%%%%%%%%%%%%%%%%%%%%%
Secondly, let us consider the case of $B_0=0$.
The equation (\ref{eq_M=0}) shows that this solution is consistent with any values of $V_1$.
Only the equation (\ref{eq_M=T}) leads to a nontrivial equation,
\begin{equation}
2\ddot{T}-(1-\dot{T}^2)(1+2\dot{T}V_0)=0.
\label{eq_B_0=0}
\end{equation}
Applying the supposed form (\ref{ansatz}) and requiring a term of order $1$ and one of order $e^{-\gamma x^0}$ to vanish separately lead to two equations given by
\begin{eqnarray}
(1+2\beta V_0)(1-\beta^2)&=&0,
\label{eq_B_0=0_0}\\
\beta-\gamma+(3\beta^2-1)V_0&=&0.
\label{eq_B_0=0_1}
\end{eqnarray}
These equations allow two solutions according to the value of $V_0$. One is given by
\begin{equation}
\beta=1, \hspace{5mm}
\gamma=2V_0+1,
\label{solution_beta=1}
\end{equation}
for $V_0>-\frac{1}{2}$.
The other solution is given by
\begin{equation}
\beta=-\frac{1}{2V_0}, \hspace{5mm}
\gamma=-\frac{1}{2}\left( 2V_0-\frac{1}{2V_0}\right),
\label{solution_beta<1}
\end{equation}
for $V_0<-\frac{1}{2}$.
The latter solution is the same as obtained in~\cite{Kluson:2003sr}.
In the case of $V_0=-\frac{1}{2}$, there is no solution with positive $\gamma$.
Note that the leading term $C$ of $\det A^{(0)}$ in equation (\ref{detA-diagonal}) vanishes for the solution (\ref{solution_beta=1}), while it does not vanish for the solution (\ref{solution_beta<1}).
This fact will turn out to cause difference of asymptotic behavior of sources for bulk fields.
\\
%%%%%%%%%%%%%%%%%%%%%%%%%%
The solution (\ref{solution_beta=1}) leads to the energy momentum tensor $T^{\mu\nu}$ expressed by the equations (\ref{T0_B0_nonzero}) (\ref{Tj_B0_nonzero}) except that $B_0$ is taken to be zero.
It is evident that the energy momentum tensor has the properties of tachyon matter.
The source for the Kalb-Ramond field $S^{\mu\nu}$ is expressed by the equations (\ref{S0_B0_nonzero}) (\ref{Sj_B0_nonzero}) except that $B_0$ is taken to be zero. This shows that there is no fundamental string charge remained in the infinite future.
\\
%%%%%%%%%%%%%%%%%%%%%%%%
The solution (\ref{solution_beta<1}) also leads to the block diagonal energy momentum tensor expressed by $T^{(0)}\oplus_{j=1}^n T^{(j)}$ with
\begin{eqnarray}
T^{(0)}&=&\tau_p\frac{1}{\sqrt{1-\beta^2}}\prod_{i=1}^n (1+B_i^2)^{1/2}
\left(
\begin{array}{cc}
1&0\\
0&-1+\beta^2
\end{array} \right) e^{\gamma x^0}+\cdots ,\\
T^{(j)}&=&-\tau_p \sqrt{1-\beta^2} \prod_{i=1,i\neq j}^n (1+B_i^2)^{1/2} (1+B_j^2)^{-1/2}
\left(
\begin{array}{cc}
1&0\\
0&1
\end{array}\right) e^{\gamma x^0}+\cdots.
\end{eqnarray}
It appears that both the energy density and the pressure diverge in the infinite future in contrast to the previous solutions.
We would simply rule out this solution.
%%%%%%%%%%%%%%%%%%%%%%%%%%%%%%%%%%%%%%
%%%%%%%%%%%%%%%%%%%%%%%%%%%%%%%%%%%%%%
\subsection{rolling tachyon with transverse electric and magnetic fields}
If an unstable D$3$-brane is considered, the analysis so far corresponds with a rolling tachyon with parallel electric field $B_0$ and magnetic field $B_1$ towards $x^1$ direction.
Subsequently, let us examine such a case that only non-vanishing components of $F_{\mu\nu}$ are $F_{01}=-F_{10}=E$ and $F_{12}=-F_{21}=B$.
This corresponds with a rolling tachyon with electric field $E$ along $x^1$ and magnetic field $B$ along $x^3$, as far as D$3$-brane is concerned.
The matrix $A_{MN}=(\eta+F)_{MN}$ is determined by $A^{(0)}\oplus 1_{p-2}$ with
\begin{equation}
A^{(0)}=\left(
\begin{array}{cccc}
1&-\dot{T}&0&0 \\
\dot{T}&-1&E&0 \\
0&-E&1&B \\
0&0&-B&1
\end{array} \right), \hspace{5mm}
\det A_{MN}=E^2-(1+B^2)(1-\dot{T}^2),
\label{non-diagonal}
\end{equation}
where $1_{(p-2)}$ denotes a rank $(p-2)$ identity matrix.
It will be found that there is a solution with non-vanishing $\det A_{MN}$.
The matrix $G^{MN}$ is determined by $G^{(0)}\oplus 1_{(p-2)}$ with
\begin{equation}
G^{(0)}=\frac{1}{\det A}\left(
\begin{array}{cccc}
-1-B^2+E^2&0&-E\dot{T}&0 \\
0&1+B^2&0&BE \\
-E\dot{T}&0&-1+\dot{T}^2&0 \\
0&BE&0&-1+E^2+\dot{T}^2
\end{array} \right). \\
\end{equation}
The matrix $\theta^{MN}$ is determined by $\theta^{(0)}\oplus 0_{(p-2)}$ with
\begin{equation}
\theta^{(0)}=\frac{1}{\det A}\left(
\begin{array}{cccc}
0&\dot{T}(1+B^2)&0&BE\dot{T} \\
-\dot{T}(1+B^2)&0&-E&0 \\
0&E&0&B(1-\dot{T}^2) \\
-BE\dot{T}&0&-B(1-\dot{T}^2)&0
\end{array} \right),
\end{equation}
where $0_{(p-2)}$ denotes a `rank' $(p-2)$ zero matrix.
The second term of the equations of motion (\ref{eom}) is found to vanish except for $M=-1,1$,
\begin{equation}
G^{MN}\partial_K F_{LN} G^{KL}= \left\{
\begin{array}{ll}
(-1-B^2+E^2)\,(1+B^2)\,\ddot{T}\,(\det A)^{-2} & \hspace*{2mm}(M=-1) \\
-E(1+B^2)\,\dot{T}\,\ddot{T}\,(\det A)^{-2} & \hspace*{2mm}(M=1).
\end{array} \right.
\end{equation}
$M=-1$ component of the equations of motion (\ref{eom}) is
\begin{equation}
(1+B^2-E^2)+2\frac{(1+B^2-E^2)(1+B^2)}{E^2-(1+B^2)(1-\dot{T}^2)}\ddot{T}+2\dot{T}\{(1+B^2)V_0+BEV_2\}=0.
\label{eom_off_M=T}
\end{equation}
$M=0$ component is
\begin{equation}
EV_1=0.
\label{eom_off_M=0}
\end{equation}
$M=1$ component is
\begin{equation}
E\dot{T}+\frac{2E(1+B^2)}{E^2-(1+B^2)(1-\dot{T}^2)}\dot{T}\ddot{T}+2EV_0+2B(1-\dot{T}^2)V_2=0.
\label{eom_off_M=1}
\end{equation}
$M=2$ component is
\begin{equation}
B(1-\dot{T}^2)V_1=0.
\end{equation}
$M=3,\cdots,p$ components do not give any non-trivial equations.
We will study a case with $E$ non-vanishing, which is possible only for $V_1=0$ according to the equation (\ref{eom_off_M=0}).
The two equations (\ref{eom_off_M=T}) and (\ref{eom_off_M=1}) give rise to
\begin{eqnarray}
2(1+B^2)\ddot{T}+(E^2-(1+B^2)(1-\dot{T}^2))(1+2\dot{T}V_0)&=&0,
\label{eq_BE=n0} \\
EV_0+BV_2&=&0.
\end{eqnarray}
The equation (\ref{eq_BE=n0}) with the asymptotic form (\ref{ansatz}) applied leads to equations for parameters $\beta$ and $\gamma$,
\begin{eqnarray}
(E^2-(1+B^2)(1-\beta^2))(1+2\beta V_0)&=&0, \\
(E^2-(1+B^2)(1-\beta^2))V_0+(1+B^2)\beta (1+2\beta V_0)-(1+B^2)\gamma &=&0.
\end{eqnarray}
These equations have two kinds of solutions. One is given by
\begin{equation}
\beta=\sqrt{1-\frac{E^2}{1+B^2}},
\hspace*{5mm}
\gamma=\sqrt{1-\frac{E^2}{1+B^2}}+2V_0 \left(1-\frac{E^2}{1+B^2}\right),
\label{BE_tachyon_matter}
\end{equation}
for $V_0>-\frac{1}{2} \sqrt{\frac{1+B^2}{1+B^2-E^2}}$. The other is
\begin{equation}
\beta=-\frac{1}{2V_0},
\hspace*{5mm}
\gamma=-\frac{1}{2} \sqrt{1-\frac{E^2}{1+B^2}} \left(2V_0 \sqrt{1-\frac{E^2}{1+B^2}}-\frac{1}{2V_0 \sqrt{1-\frac{E^2}{1+B^2}}} \right),
\label{BE_not_tachyon_matter}
\end{equation}
for $V_0<-\frac{1}{2} \sqrt{\frac{1+B^2}{1+B^2-E^2}}$.
The latter solution (\ref{BE_not_tachyon_matter}) leads to the singular behavior of energy momentum tensor, thus we would simply rule out this solution.
The former solution (\ref{BE_tachyon_matter}) leads to energy momentum tensor with relevant components determined by
\begin{eqnarray}
T^{\mu\nu}&=&\frac{\tau_p}{\sqrt{2\beta K(1+B^2)}}
\left(
\begin{array}{ccc}
1+B^2&0&BE \\
0&\frac{-E^2}{1+B^2}&0 \\
BE&0&\frac{B^2E^2}{1+B^2}
\end{array} \right) e^{-V_i x^i+\frac{BEV_2}{1+B^2} x^0}
+\cdots , \\
T^{ab}&=& -\tau_p \sqrt{2\beta K(1+B^2)}\, \delta^{ab} e^{-V_i x^i+\left(\frac{BEV_2}{1+B^2}-\gamma \right)x^0}+\cdots,
\end{eqnarray}
with $\mu,\nu=0,1,2$,\hspace{2mm} $a,b=3,\cdots,p$,\hspace{2mm}$i=2,\cdots,p$.
The source for the Kalb-Ramond field is determined by relevant components,
\begin{equation}
S^{\mu\nu}=-\frac{\tau_p}{\sqrt{2\beta K(1+B^2)}}
\left(
\begin{array}{ccc}
0&-E&0 \\
E&0&\frac{BE^2}{1+B^2} \\
0&-\frac{BE^2}{1+B^2}&0
\end{array} \right) e^{-V_ix^i+\frac{BEV_2}{1+B^2} x^0}
+\cdots ,
\label{S01}
\end{equation}
with $\mu,\nu=0,1,2$ and other components zero.
Note that momentum density to $x^2$ direction is finite, hence the whole system moves towards $x^2$ direction with velocity determined by $T^{02}/T^{00}\;=\frac{BE}{1+B^2}$.
This fact shows that $T^{\mu\nu}$ and $S^{\mu\nu}$ are not changed in the rest frame, while pressures to direction $x^a$, $T^{aa}$, goes to vanish in the infinite future. This solution is still qualified to represent tachyon matter.
In the presence of spacelike linear dilaton background, profiles of the energy-momentum tensor and the source of the Kalb-Ramond field vary towards a certain spatial direction at a specific time in a given coordinate frame. Therefore we note that the rest frame should be used to examine the profiles.

The value of dilaton changes in the rest frame according to
\begin{equation}
\Phi= V_0\frac{1+B^2-E^2}{\sqrt{(1+B^2)^2-B^2E^2}}\, {x'}^0,
\end{equation}
where ${x'}^0$ denotes the time coordinate in the rest frame.
Since the electric field is allowed to take a value subject to $(1+B^2-E^2)>0$, the direction of effective coupling evolution for the tachyon matter solution depends only on the sign of $V_0$, which is not restricted in this analysis.

%%%%%%%%%%%%%%%%%%%%%%%%
If the fluctuation modes around the tachyon matter solution can propagate, it would give significant implications for the fundamental string formation.
Then, according to~\cite{Gibbons:2002tv}, let us study the fluctuation modes defined by
\begin{equation}
T=T^{\tiny\mbox{tm}}+t, \hspace{5mm}
A_{\mu}=A^{\tiny\mbox{tm}}_{\mu}+a_{\mu} \hspace{3mm} (\mu=0,1,\cdots,p),
\label{def-fluctuation}
\end{equation}
where $T^{\tiny\mbox{tm}}$ and $A^{\tiny\mbox{tm}}_{\mu}$ denote the tachyon matter solution (\ref{BE_tachyon_matter}), and fluctuation modes $t$ and $a_{\mu}$ may depend on $x^{\mu}$. For simplicity, we use a gauge $a_0=0$.
The tachyon matter solution gives rise to vanishing $\det (\eta+F)_{MN}$ up to negligible terms of order $e^{-\gamma x^0}$. Accordingly, it is sufficient for fluctuation analysis to examine equations that come from the second term of the equations of motion,
\begin{equation}
\partial_K F_{LN} G^{KL}=0.
\label{fluctuation}
\end{equation}
up to terms of order $e^{-\gamma x^0}$ and second order terms for fluctuation modes.\\
$N=-1$ component of this equation is
\begin{equation}
(1+B^2)\partial_0^2 t-\frac{E^2}{1+B^2}\partial_1^2 t+\frac{B^2E^2}{1+B^2}\partial_2^2 t+2BE\partial_0\partial_2 t=0.
\label{non-diagonal-fluctuation-eqT}
\end{equation}
$N=0$ component is
\begin{equation}
E\partial_0 \hat{t}=0, \hspace{5mm}
\hat{t}\equiv \beta \partial_1 t+\frac{E}{1+B^2} \partial_1 a_1-B\partial_0 a_2-\frac{B^2E}{1+B^2}\partial_2 a_2.
\label{non-diagonal-fluctuation-eq0}
\end{equation}
$N=\mu \hspace{2mm}(\mu=1,\cdots,p)$ components are
\begin{equation}
(1+B^2)\partial_0^2 a_{\mu}-\frac{E^2}{1+B^2}\partial_1^2 a_{\mu}+\frac{B^2E^2}{1+B^2}\partial_2^2 a_{\mu}+2BE\partial_0\partial_2 a_{\mu}+E\partial_{\mu} \hat{t}=0.
\label{non-diagonal-fluctuation-eq-mu}
\end{equation}
We will study a case with non-vanishing $E$.
It follows from the equation (\ref{non-diagonal-fluctuation-eq0}) that $\hat{t}$
is not dynamical. Moreover it can be set to be zero by using a spatial gauge transformation that still remains in the gauge $a_0=0$.
The equations (\ref{non-diagonal-fluctuation-eq-mu}) subject to $\hat{t}=0$ give rise to the same kind of differential equation as (\ref{non-diagonal-fluctuation-eqT}), which can be expressed as
\begin{equation}
\left(\partial_0+\frac{E}{1+B^2}\partial_1+\frac{BE}{1+B^2}\partial_2\right)\left(\partial_0-\frac{E}{1+B^2}\partial_1+\frac{BE}{1+B^2}\partial_2\right)t=0.
\end{equation}
This indicates that there are $p$ fluctuation modes that can propagate with velocity $\frac{BE}{1+B^2}$ to the $x^2$ direction and $\pm\frac{E}{1+B^2}$ to the $x^1$ direction, though the world volume is extended to a $p$-dimensional space.
The fact that the whole system moves towards the $x^2$ direction with velocity $\frac{BE}{1+B^2}$ means that an intrinsic propagation of fluctuation modes is towards $x^1$ direction, that is, the direction of electric flux.

The constant electric flux is generically identified with the conserved fundamental string charge. In fact, the equation (\ref{S01}) shows a presence of fundamental string charge along the $x^1$ direction.
Accordingly, if a spatially inhomogeneous tachyon condensation is considered, fundamental string would be formed towards the direction of electric flux.
We note that the finite electric flux is not admissible towards directions with finite linear dilaton. Thus, it appears that fundamental string is not formed towards the finite valued linear dilaton.

The agreement of the possible direction of the intrinsic fluctuation propagation with the constant electric flux suggests that the propagating fluctuation modes come from the fundamental strings stretched along electric flux.
This assumption is consistent with the fact that the nature of propagation of fluctuation modes is not changed by the presence of linear dilaton background, since a fundamental string itself is not affected by the value of effective coupling.
Incidentally the propagation speed of fluctuation modes vanishes in the $E\to 0$ limit, and the unstable D-brane does not need D1-brane charge at all. The propagating fluctuation mode should come from a small wave on the fundamental string itself, and not from open string on D1-brane with fundamental string resolved on it.

%%%%%%%%%%%%%%%%%%%%%%%%%%%%%%
%%%%%%%%%%%%%%%%%%%%%%%%%%%%%%
\section{Conclusion}
We have considered the rolling tachyon in the linear dilaton background by using the effective field theory with gauge field.
Several asymptotic solutions equipped with or without constant electromagnetic fields were obtained in linear dilaton background.
We explicitly constructed a solution to be identified with tachyon matter in linear dilaton background. Moreover tachyon matter solutions with constant electromagnetic field in linear dilaton background were also obtained.
We showed that tachyon matter solutions are obtained only for $V_0$ larger than certain constants, which may depend on the profile of electromagnetic field.
The existence of the lower bounds was not seen in the analysis by using the timelike boundary Liouville theory~\cite{Karczmarek:2003xm}.
This discrepancy implies that the effective action (\ref{action}) does not correctly describe a certain parameter region.
Indeed, the effective action is not shown to be valid in a region where the dilaton is large. We would insist that the result for $V_0=0$ at least is to be reliable in a region where the dilaton is not so large. Then the apparent discrepancy disappears.

It turned out that the direction of time evolution of effective coupling is determined by the sign of $V_0$ and the profile of electromagnetic field enters into the speed of evolution.
The nature of fluctuation modes such that the intrinsic propagation is towards the electric flux and the propagation speed is proportional to the electric flux was shown to be unchanged in the presence of linear dilaton background.
This result is consistent with the assumption such that the propagating fluctuation modes come from long stretched fundamental strings.
%%%%%%%%%%%%%%%%%%%%%%%%%%%%%%%%%%%%%%%%%%%%%%%
%%%%%%%%%%%%%%%%%%%%%%%%%%%%%%%%%%%%%%%%%%%%%%%
\section*{Acknowledgements}
The author would like to thank to K. Ito for useful comments.
He would like to thank to H.~Kunitomo and S.~Sugimoto for helpful discussions.
%%%%%%%%%%%%%%%%%%%%%%%%%%%%%%%%%%%%%%%%%%%%%%%
%%%%%%%%%%%%%%%%%%%%%%%%%%%%%%%%%%%%%%%%%%%%%%%

\end{document}